\input harvmac
\input epsf
\epsfverbosetrue

\def\p{\partial}

\def\b0{\bar{0}}
\def\b4{\bar{4}}
\Title{EFI-98-05}{\vbox{\centerline{Comments on Supersymmetric Yang-Mills
Theory }
\vskip12pt
\centerline{on a Noncommutative Torus}
}}
\vskip20pt
\centerline{Miao Li}
\bigskip
\centerline{\it Enrico Fermi Institute}
\centerline{\it University of Chicago}
\centerline{\it 5640 Ellis Avenue, Chicago, IL 60637, USA}

\bigskip
\centerline{\it }
\centerline{\it }
\centerline{\it }
\bigskip
D0-brane theory on a torus with a nonvanishing $B$ field is embedded into
a string theory in the weak coupling limit. It is shown that the usual
supersymmetric Yang-Mills theory on a noncommutative torus can not be
the whole story. The Born-Infeld action survives the noncommutative torus
limit.

\Date{Feb. 1998}

\nref\cds{A. Connes, M.R. Douglas and A. Schwarz, hep-th/9711162.}
\nref\dh{M.R. Douglas and C. Hull, hep-th/97111165.}
\nref\lr{R. Leigh and M. Rozali, hep-th/9712168.}
\nref\hw{P.-M Ho, Y.-Y. Wu and Y.-S. Wu, hep-th/9712201;
P.-M. Ho and Y.-S. Wu, hep-th/9801147.}
\nref\rc{R. Casalbuoni, hep-th/9801170.}
\nref\seiberg{N. Seiberg, hep-th/9710009; A. Sen, hep-th/9709220.}
\nref\suss{L. Susskind, hep-th/9704080.}
\nref\tseytlin{A. Tseytlin, hep-th/9701125.}
\nref\bs{T. Banks and N. Seiberg, hep-th/9702187.}
\nref\wati{W. Taylor, hep-th/9611042.}

In a pair of interesting papers by Douglas and collaborators \refs{
\cds, \dh} (for further development see \refs{\lr, \hw, \rc}),
it was pointed out that the super Yang-Mills theory on a noncommutative
torus is naturally related to compactification of matrix theory
on a dual torus with a constant $C^{(3)}$ field. This field has a
tangent index along the longitudinal direction as well as 
two indices along the compact torus. These two indices meet
the minimal requirement of noncommutativeness. Thus
the geometric correspondence
to a noncommutative torus is quite natural, and the vertices
arising from the noncommutativeness can be derived in string theory.

The field theory is superficially nonrenormalizable, since the 
action involves infinitely many high derivative terms. However these
terms sum to an exponential form, and this becomes a damping factor
at higher energies. It may well be that this theory is a  
well-defined quantum field theory on a two dimensional or three 
dimensional torus.
Our aim in this note is not to attempt a front attack on this
renormalization problem. We shall try to embed these theories into
string theory, and ask whether there is a proper limit in which
the theory is decoupled from string theory.

We shall argue that on a two dimensional torus, like the $C=0$ case,
the SYM can be embedded into a weakly coupled string theory.
Actually, the string coupling constant for the fixed dimensionless
Yang-Mills couplings tends to zero faster than in the case when
$C=0$. This indicates that the SYM on the noncommutative torus
is indeed renormalizable. The fact that the decoupling works better
than on a usual torus might have some root in  an intrinsic
property of SYM on a noncommutative torus: The nonlocal vertices
have damping effects at high energies. The new result of our analysis
is that in addition to the higher derivative terms originating
from the noncommutative torus, there are higher derivative terms
from the Born-Infeld action which are also important. Actually,
when the compactification scales are smaller than the Planck scale,
the  Born-Infeld dominates.

Seiberg argued that the DLCQ of matrix theory on a finite light-like
circle can be obtained by infinitely boosting a small space-like
circle. The M theory on the space-like circle is a weakly coupled
IIA string. Its dual on a two torus is again a weakly coupled string.
In the three torus case, the dual string coupling is fixed.
In this way one argues that indeed the 2D and 3D SYM decouple from
the corresponding string theory in the limit $l_s\rightarrow 0$.
Applying the infinite boost argument, we find the following:
On a two torus, the string coupling constant still goes to
zero faster than the $C=0$ case. The Born-Infeld action survives
in the decoupling limit.

Consider D0-branes on two torus $T^2$
of size $(R_1, R_2)$. Assume this torus be a rectangular one.
A slanted torus introduces no new novelty. A constant $B_{12}=B$ is 
switched
on. Our normalization for the $B$ field is such that it is dimensionless
and the coupling to the string world sheet is
\eqn\bcoup{\int B_{\mu\nu}dx^\mu\wedge dx^\nu ,}
where we always use the coordinates with the period $2\pi$. Thus,
on the two torus the metric is $ds^2=(R_1dx_1)^2+(R_2dx_2)^2$.
The two complex moduli are given by
\eqn\moduli{\tau ={R_1\over R_2}i, \quad \rho =B+l_{s}^{-2}R_1R_2 i.}
Following Douglas and Hull, we switch to the T-dual torus obtained by
performing T-duality along $R_2$. This amounts to exchanging the two 
moduli $\tau$ and $\rho$. In this dual picture, the new field
$B' =0$, and the two new radii are
\eqn\newr{R_1'={1\over R_2}\sqrt{l_s^4 B^2+R_1^2R_2^2}, \quad
R_2'={l_s^2\over R_2}.}
An original D0-brane is transformed to a D-string wrapped in the 
$x_2'$ direction. The new torus is a slanted one with the angle
$\theta$ determined by $\sin\theta =R_1R_2/\sqrt{l_s^4B^2+R_1^2R_1^2}$.
The shortest string stretched between a D-string and its nearest
image has a mass
\eqn\mass{M=l_s^{-2}R_1'\sin\theta =l_s^{-2}R_1.}
\bigskip
{\vbox{{\epsfxsize=2.5in
        \nobreak
    \centerline{\epsfbox{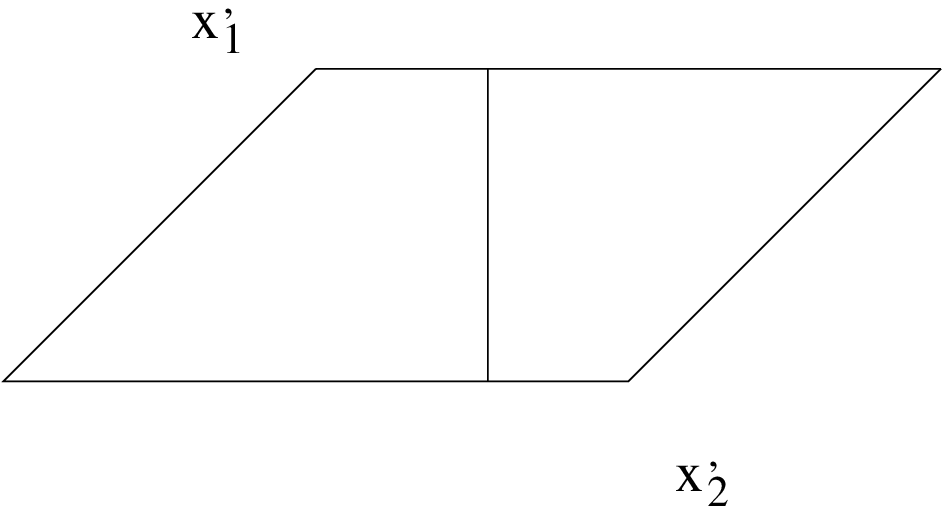}}
        \nobreak\bigskip
    {\raggedright\it \vbox{
{\bf Figure 1.}
{\it The shortest string stretched between a D-string and its image.}
 }}}}}
\bigskip

There is a displacement between the two ends in the new coordinate
$x_2'$, $\delta x_2'=2\pi Bl_s^{2}/R_2$. This is the origin of
the nonlocality of the field theory describing winding modes on the
original two torus. In a vertex involving three open string fields
there is an insertion of operator
\eqn\vertex{\exp (i2\pi B(\p_1^1\p_2^2-\p_1^2\p_2^1)),}
where we used dual coordinates with period $2\pi$ which parametrize
winding numbers. $\p_i^a$ denotes $\p_i$ acting on the $a$-th field.
\vskip18pt
{\vbox{{\epsfxsize=3in
        \nobreak
    \centerline{\epsfbox{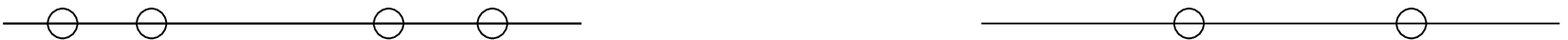}}
        \nobreak\bigskip
    {\raggedright\it \vbox{
{\bf Figure 2.}
{\it An open string appears on the D-string as a dipole. Two dipoles
join to form a larger dipole, this is the origin of nonlocality.}
 }}}}}
\bigskip

The stretched string described above can be interpreted as a winding mode
stretched between a D0-brane and its nearest image in the $x_1$ 
direction, on the original torus. A more general formula for the mass for 
a winding open string is given by
\eqn\omass{M=l_s^{-2}\sqrt{(mR_1)^2+(nR_2)^2}.}
Ignoring the D0-brane and its image on which this string ends, it
can be regarded as a closed string of winding numbers $m$, $n$.
However, one should distinguish between this open string mode and the
corresponding closed string mode. Indeed, the mass formula for the
latter is
\eqn\cmass{M={\sqrt{l_s^4B^2+R_1^2R_2^2}\over R_1R_2}l_s^{-2}
\sqrt{(mR_1)^2+(nR_2)^2}.}
This mass is larger than the mass of the open string state with
the same winding numbers.
It depends on the background field $B$. The reason for the independence
of the mass on $B$ of an open string is simple: The world sheet coupling
$\int B$ vanishes for an open string upon imposing the Dirichlet
boundary conditions. This fact meshes well with the mass formula we
obtained on the dual torus $(x_1', x_2')$, eq.\mass. On the other hand, the 
world sheet coupling of $B$ to a closed winding string contributes to the
quantization of the spectrum. Although the spectrum of open string
decouples from $B$, the interaction does not. The standard way of
computing interaction is by insertions of the open string vertex 
operators on the boundary of the world sheet. Now an anomaly will
arise from the jump of boundary conditions crossing a vertex
operator.

To decouple these winding open string modes from the open string modes
with oscillators
as well as closed string modes with oscillators, the necessary condition
is $R_i\ll l_s$ such that the energy of a stretched open string is much
smaller than the string scale. One would expect that in this limit
these modes are also decoupled from the closed string massless states.
This is simply because the coupling between these two sets are not
set by $g_s$, but by $\kappa =g_sl_s^4$. However, we shall use 
$g_s$ on the dual torus obtained by performing T-duality along both
directions $x_i$. The new complex moduli associated to this dual torus
are given by
$$\tilde{\tau}=-{1\over \tau}, \quad \tilde{\rho}=-{1\over\rho}.$$
The new torus is still a rectangular one with nonvanishing
$B$ field. The real parameters are read off from the above
equations
\eqn\newmu{\eqalign{\Sigma_1&={l_s^2R_2\over \sqrt{l_s^4B^2+R_1^2R_2^2}},\cr
\Sigma_2 &={l_s^2 R_1\over\sqrt{l_s^4B^2+R_1^2R_2^2}},\cr
\tilde{B} &=-{l_s^4B\over l_s^4B^2+R_1^2R_2^2},}}
Although the new radii depend on $B$, the mass of a momentum mode
is independent of $B$. This is because it corresponds to a winding
mode on the original torus. Another way to see this is by a direct
computation. This time $\tilde{B}$ is not decoupled in the world
sheet action, since the relevant boundary condition is Neumann.
In the low energy limit $R_i/l_s\rightarrow 0$, we can ignore
the second term in $l_s^4B^2+R_1^2R_2^2$ for a fixed $B$. Thus,
the new radii are given by $\Sigma_1=R_2/B$,
$\Sigma_2=R_1/B$. These are much different from the formulas
when $B=0$. For a background $B\sim 1$, the size of the dual torus
is the same order of the size of the original torus. In the limit
we are interested in, both are very small.
At this point it is interesting to note that
it is impossible for an open string momentum mode to decay into
closed string momentum modes. This is simply due to the fact
that the energy scale of the former is $R_il_s^{-2}$, much smaller than
the energy scale $B/R_i$ of the latter. 

Ignoring the high derivative terms introduced by noncommutativeness
of the dual torus, the low energy theory is a $2+1$ SYM theory.
This theory is well-understood and is the low energy world-volume
theory of D2-branes. Naively, one has a paradox here. If the 
$2+1$ Yang-Mills is the theory of D2-branes wrapped on the dual
torus, why is not the spectrum determined by $\Sigma_i$ which
depend on $B$, while is determined by the sizes $l_s^2/R_i$ as
if there is no $B$ field? Furthermore, what is the Yang-Mills
coupling constant, is it still given by $\tilde{g}_sl_s^{-1}$?
A careful analysis of the low energy action will help to answer
these questions. 
				
For convenience, we set $l_s=1$. The low energy
condition becomes $R_i\ll 1$. The standard Born-Infeld action reads
\eqn\bi{S=-{1\over\tilde{g}_s}\int d^3x\hbox{det}^{1/2}\left(G_{\mu\nu}+
F_{\mu\nu}-B_{\mu\nu}\right),}
where for simplicity we dropped the tilde symbol. We also suppressed
terms associated to scalars.
If we expand the above action in the usual fashion, assuming $G_{\mu\nu}$
dominates other two terms in the determinant, we would obtain the usual
low energy Yang-Mills action with $g_{YM}^2=\tilde{g}_sl_s^{-1}$. 
In the case of interest, $G_{\mu\nu}$ are given by $\Sigma_i^2$ which
are much smaller than the $B$ field, thus we shall expand the determinant
in a different way. The determinant is
$$\hbox{det}^{1/2}\left(G_{\mu\nu}+F_{\mu\nu}
-B_{\mu\nu}\right)=\alpha [1-{1\over\alpha^2}(2\tilde{B}F_{12}-F_{12}^2
+\Sigma_2^2F_{01}^2+\Sigma_1^2F_{02}^2)]^{1/2},$$
where $\alpha^2=\tilde{B}^2+\Sigma_1^2\Sigma_2^2$. We see that the
terms $F_{0i}^2$ are weighted differently than $F_{12}$. The weights of
the former are much smaller. As we shall see in a moment, for an on-shell
field configuration, $F_{0i}^2\sim \Sigma^2 F_{12}^2$. Thus, we can ignore
higher orders in $F_{0i}^2$ in the expansion of the determinant, except 
for the quadratic ones which are needed for having interesting dynamics.

Thus, the appropriate low energy action is
\eqn\effac{S={1\over 2\tilde{g}_s\alpha}\int\left(\Sigma_2^2F_{01}^2+
\Sigma_1^2F_{02}^2)-2\alpha^2\sqrt{1+{1\over \alpha^2}(F_{12}^2-2\tilde{B}
F_{12})}\right).}
Since $\alpha^2\sim \tilde{B}^2\sim 1$, naively one expects that the 
quadratic term in $F_{12}$ has a coefficient of order $1$. This is
incorrect, for the second term in the expansion of the square root
almost cancels the first (we drop the linear term in $F_{12}$, it
is a total derivative). So to the quadratic order,
\eqn\quadra{S= {1\over 2\tilde{g}_s\alpha}\int \left(\Sigma_2^2F_{01}^2+
\Sigma_1^2F_{02}^2-({\Sigma_1\Sigma_2\over \alpha})^2F_{12}^2\right).}
An on-shell state will have, for instance $\Sigma_2^2F_{01}^2\sim
(\Sigma_1\Sigma_2/\alpha)^2F_{12}^2$. This justifies eq.\effac.
Now use $\alpha\sim \tilde{B}\sim 1/B$ and $\Sigma_i\sim \epsilon_{ij}
R_j/B$, and do rescaling $x^i\rightarrow R_ix_i$, $A_i\rightarrow
1/R_i A_i$, the Yang-Mills action is put into the standard form
$F_{\mu\nu}^2$ with the coupling $g^2_{YM}=\tilde{g}_sB/(R_1R_2)$.
The fact that the Yang-Mills has the standard form in the new coordinates
system implies that the dispersion relation of the spectrum is exactly
the same as we argued for before, eq.\omass. A typical momentum mode has 
a mass
$R_i$, much smaller than the scale $1/\Sigma_i$. The Yang-Mills coupling
is quite different from $\tilde{g}_s (l_s^{-1})$. Indeed, use the
T-duality relation $\tilde{g}_s=g_s/B$, $g_{YM}^2=g_s/(R_1R_2)$, the
same as one might expect from the T-duality relation for $B=0$. 
Since the energy gap is $R_i$,
the dimensionless couplings are $g^2_{YM}/R_i=\tilde{g}_sB/(R_i)^3$.
For intermediate such couplings, we require $\tilde{g}_s\sim 
(R_i)^3/B\ll 1/B$. The string theory is weakly coupled. When $B=0$,
$g^2_{YM}=\tilde{g}_s$ and the dimensionless couplings are given
by $\tilde{g}_s/R_i$. For intermediate couplings we have
$\tilde{g}_s\sim R_i \ll 1$. Compared to the case $B\sim 1$, $\tilde{g}_s
$ goes to zero more slowly.

It is important to realize that we recover the correct physics from
the Born-Infeld action only when we expand the square root of the
determinant to the second order. In the $B=0$ case, the low
energy physics is reproduced by taking only the first order terms.
Indeed, there are higher order terms in $F_{12}$ and these terms can
not be ignored in the low energy physics. We now show that for a single
quanta, the terms in the square root of eq.\effac\ are comparable to 1.
Use the normalization in \effac, a single quanta of energy $R$
(assuming $R_1\sim R_2\sim R$) corresponds to the field strength
$F_{12}^2\sim \tilde{g}_sBR^{-3}$. This gives 
\eqn\nonl{{1\over\alpha^2}F_{12}^2=B^2F_{12}^2 \sim B^3(\tilde{g}_s R^{-3}),
\quad {\tilde{B}\over\alpha^2}F_{12}=BF_{12}=B^{3/2}(\tilde{g}_s 
R^{-3})^{1/2}.}
For the intermediate Yang-Mills coupling, $\tilde{g}_s\sim R^3/B$.
So the above terms are comparable to $1$ if $B\sim 1$, and we have
to use the effective action \effac, or better, the full Born-Infeld 
action \bi.

It then appears that there is a large correction to the mass of a 
momentum mode in the gauge theory. We expect that this correction 
vanishes for some BPS states. 
A state with constant $F_{12}$ represents a bound state of  D0-branes
and a D2-brane, and indeed the presence of $\tilde{B}$ causes
a correction to the mass formula, and the correction is large
in the regime of noncommutative torus. We have not discussed the
scalar fields. It is not hard to see that the dispersion relation is
what was expected, and the correction of the Born-Infeld
action to their kinetic energy is not large.
For a non-BPS process involving $F_{12}$, the interaction
determined by the Born-Infeld action is large. It is larger than the
kinetic energy, and also larger than the interaction energy caused
by the higher derivative terms coming from the noncommutativeness.
For $B\sim 1$, the latter is comparable to the kinetic energy.

Recently Sen and Seiberg proposed a systematical approach to matrix
theory on a torus \seiberg. Seiberg's argument involves boosting along a 
small spatial circle to get a near light-like circle of radius $R$.
The resulting matrix theory is the DLCQ version proposed by Susskind
\suss. We now employ Seiberg's boost argument to determine how important
the Born-Infeld action is in matrix theory. Let the Planck mass scale be
$M_P$ and the radius of the light-like circle be $R$. The corresponding
scales on the small spatial circle are $m_P$ and $R_s$. Consider 
compactification on a two torus of radii $R_i$. The corresponding radii
in the other theory with a spatial M circle are denoted by $r_i$.
The scales in the two theories are related through
\eqn\rela{R_sm_P^2= RM_P^2, \quad r_im_P=R_iM_P.}
Consequently
\eqn\para{\eqalign{g_s &=R_s^{3/4}(RM_P^2)^{3/4}, 
\quad l_s^2=R_s^{1/2}(RM_P^2)^{-3/2},\cr
r_i&=({R_s\over R})^{1/2}R_i.}}
The matrix theory is obtained as the D0-brane theory in the limit 
$R_s\rightarrow 0$.
Following \cds, we switch on a $C$ field $C_{-12}\ne 0$. This is the
field in the DLCQ M theory. Let the corresponding field in the M theory
with a small spatial circle be denoted by $\tilde{C}_{-12}$. There
must be a relation
\eqn\match{M_P^3RR_1R_2C_{-12}=m_p^3R_sr_1r_2\tilde{C}_{-12}.}
The $B$ field is given by $B=l_s^{-2}r_1r_2\tilde{C}_{-12}$. Using the
above relation $B=M_P^3RR_1R_2C_{-12}$. We see that $B$ is fixed in the limit
$R_s\rightarrow 0$, and indeed there is a noncommutative torus.

The Yang-Mills coupling on the dual torus is, according to the previous
analysis,  $g_{YM}^2=\tilde{g}_sl_sB/(r_1r_2)=g_sl_s/(r_1r_2)
=R/(R_1R_2)$. It is fixed in the limit $R_s\rightarrow 0$. Note that
$\tilde{g}_s$ goes to zero as $R_s\rightarrow 0$, and the D2-brane theory
is embedded into a weakly coupled string theory.

We have seen that the important terms in the Born-Infeld action for
a momentum quanta are of the order given in \nonl. The $F_{12}^2$ term
is estimated to be $B^3\tilde{g}_sl_s^3r^{-3}_i=B^2g_sl_s^3r_i^{-3}$, where
we restored a factor associated to $l_s$. Using relations \para,
this is $B^2(M_PR_i)^{-3}$. This is finite in the limit $R_s\rightarrow
0$, and is small only when the compactification scale $R_i$ is much
larger than the Planck scale. M theory compactified on a two torus
correspond to IIA/IIB string theory. When  both scales
of the two torus are large, this is the strong coupling string limit.
(One may employ the $SL(2,Z)$ duality to go to a weak coupling limit.)
If one of the circles, say $R_1$, is much smaller than the
Planck length, this is a weakly coupled IIA string, and the Born-Infeld
action can not be ignored.

Generalization to include multiple D0-branes poses a serious problem.
The full non-abelian Born-Infeld action is not known, despite an
interesting proposal \tseytlin. In any case, inclusion of Born-Infeld
adds to the problem of renormalization. 

One way to get around the problem associated with the Born-Infeld action
is to go to the large N limit and try to decouple momentum modes.
The momentum modes represent longitudinal objects in matrix theory, and
in the decompactification limit of the longitudinal direction, these
are much too heavy to play a role in the dynamics. In the large N limit,
a toron, for instance, is described by a small $F_{12}$ proportional
to $1/N$, and the Born-Infeld action is not important for such a small
field strength. Thus a transverse membrane in matrix theory can be described
by the usual matrix theory action. It must be noted that, however, the 
light-cone energy of the bound state of a transverse membrane and 
N D0-branes gets corrected by a constant term $B/R$. This is simply
the statement that a transverse membrane in the background $B=RR_1R_2C$
induces a D0-brane charge $B$. This is achieved in \cds\ by adding
a term $B\int F_{12}$ to the action. Here this term is a consequence of 
the expansion of the Born-Infeld action \bi. In the matrix string
context, the momentum modes are important, but now in a twisted sector.
$F_{12}$ can be made arbitrarily small again, if the length of the
twisted sector is long enough. It is an interesting question whether
the gauge field can be dualized to a scalar on a noncommutative torus.
This dualization plays a crucial role in recovering the eighth scalar
in the light-cone string theory \bs.

It is not clear to us whether the $B$ moduli  should play a role in the 
decompactification limit of the longitudinal direction, since to hold
$B$ fixed, $C_{-12}\rightarrow 1/R$. There is no doubt that for finite $R$,
one can always switch on a nonvanishing $C$ field. In such a case
the inclusion of momentum modes of the gauge field necessitates the use
of the Born-Infeld action. It is therefore desirable to know the whole
action which includes all the relevant higher derivative terms.
For the time being, we do not know how to do this. The best way to see
a vertex associated with the noncommutative torus is to work with
D-strings obtained after T-dualizing one circle. The best way to
obtain the Born-Infeld action is to T-dualize both circles and work
with a constant background of $F_{\mu\nu}$. Now, since the Born-Infeld
involves vertices higher order in the open string field, one might follow
Douglas and Hull to derive the general form of a differential operator
such as \vertex. For example, consider a vertex involving four fields.
Assume three fields representing three open strings join to form a forth
open string. The differential operator is a product of three operators like
\vertex, each is concerned with a pair of fields out of the three fields.
In general, define the $*$ product for two fields as follows
\eqn\star{\phi_1(x_1,x_2)*\phi_2(x_1,x_2)=e^{i2\pi B(\p_1^1\p_2^2
-\p_1^2\p_2^1)}\phi_1\phi_2,}
the three vertex can be written as
\eqn\three{\int\phi_3(\phi_1*\phi_2).}
A four vertex can be written as
\eqn\four{\int \phi_4((\phi_1*\phi_2)*\phi_3),}
and another possible form
\eqn\foura{\int (\phi_1*\phi_2)(\phi_3*\phi_4),}
etc. The $*$ product is associative, not commutative. This reflects
the nature of open string interaction. Of course an actual vertex may
contain additional differential operators.

Higher order terms contained in the BI action are weighted by $B$ as
well as $g_s$. It is difficult to see how these terms could arise in
the D-string picture, since there is a slanted torus without a $B$
background. One would say that the calculation of interaction is
identical to the one on a rectangular torus without $B$, except for
insertions of differential operators \vertex.  Is it possible that further
terms are needed to cure some problem caused by nonlocality?

Finally, we do not see how to derive higher vertices from the matrix
lagrangian, except those contained in the Yang-Mills action, 
using the procedure of \wati. On general grounds, the IMF Hamiltonian
for a given background is not necessarily derivable from the IMF
Hamiltonian of another background, due to the subtlety brought about
by integrating out zero modes. Here might be a simple example demonstrating
this point.

\noindent{\bf Acknowledgments} 
We would like to thank M. Douglas , E. Martinec and Y.-S. Wu for comments.
This work was supported by DOE grant DE-FG02-90ER-40560 and NSF grant
PHY 91-23780.

\listrefs
\end